\documentclass[journal]{IEEEtran}

\usepackage{graphicx}
\usepackage{wrapfig}
\usepackage{amsmath}
\usepackage{subfig}
\usepackage{esvect}
\usepackage[utf8]{inputenc}
\usepackage{color}
\usepackage[square,sort,comma,numbers]{natbib}
\usepackage{amssymb}
\renewcommand{\arraystretch}{1.5}
\usepackage[numbers]{natbib}
\usepackage{empheq} 
\usepackage{amsmath}
\usepackage{commath}
\usepackage{indentfirst}
\usepackage{amsfonts}
\usepackage{xcolor}

\DeclareMathOperator*{\argmin}{argmin}

   \usepackage{tabu}
\begin{document}

\title{Data-driven Seismic Waveform Inversion:\\ A Study on the Robustness and Generalization}

\author{Zhongping Zhang$^{1}$ and Youzuo Lin$^{1, *}$,
\thanks{\textbf{1}: the Earth and Environmental Sciences, Los Alamos National Laboratory, Los Alamos,
NM, 87544 USA.}
\thanks{Correspondence to: Y. Lin, ylin@lanl.gov.}}


\maketitle

\begin{abstract}
Acoustic- and elastic-waveform inversion is an important and widely used method to reconstruct subsurface velocity image. Waveform inversion is a typical non-linear and ill-posed inverse problem. Existing physics-driven computational methods for solving waveform inversion suffer from the cycle skipping and local minima issues, and not to mention solving waveform inversion is computationally expensive. In recent years, data-driven methods become a promising way to solve the waveform inversion problem. However, most deep learning frameworks suffer from generalization and over-fitting issue. In this paper, we developed a real-time data-driven technique and we call it VelocityGAN, to accurately reconstruct subsurface velocities. Our VelocityGAN is built on a generative adversarial network (GAN) and trained end-to-end to learn a mapping function from the raw seismic waveform data to the velocity image. Different from other encoder-decoder based data-driven seismic waveform inversion approaches, our VelocityGAN learns regularization from data and further impose the regularization to the generator so that inversion accuracy is improved. We further develop a transfer learning strategy based on VelocityGAN to alleviate the generalization issue. A series of experiments are conducted on the synthetic seismic reflection data to evaluate the effectiveness, efficiency, and generalization of VelocityGAN. We not only compare it with existing physics-driven approaches and data-driven frameworks but also conduct several transfer learning experiments. The experiment results show that VelocityGAN achieves state-of-the-art performance among the baselines and can improve the generalization results to some extent.

\end{abstract}

\begin{IEEEkeywords}
full-waveform inversion, data-driven method, condition adversarial networks, transfer learning
\end{IEEEkeywords}

\IEEEpeerreviewmaketitle

\section{Introduction}

\begin{figure}[th]
\begin{center}
    \includegraphics[width=0.75\linewidth]{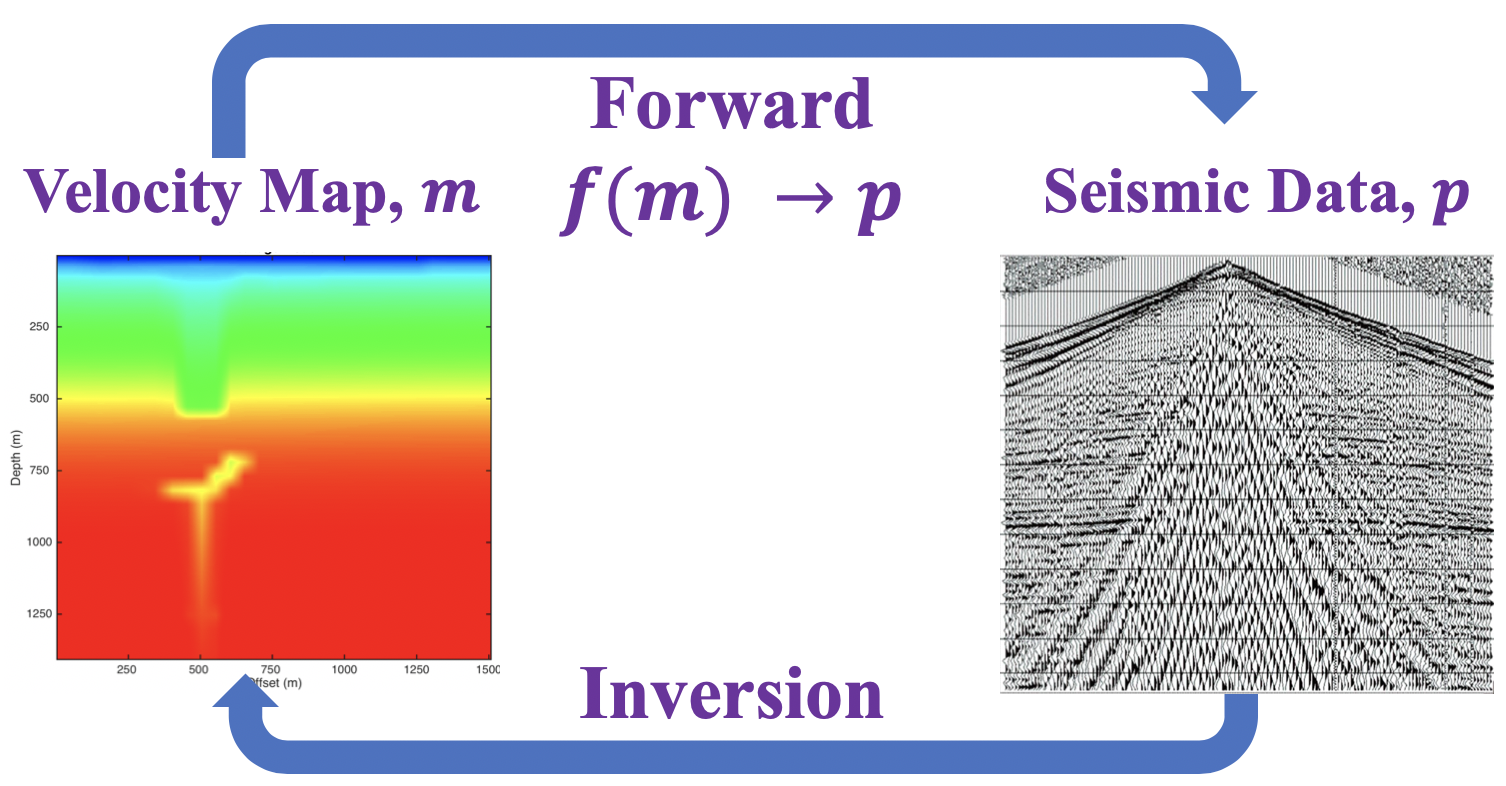}
\end{center}
   \caption{The forward model (the physics) takes a map of seismic wave velocity as input and by simulating the physics of wave propagation in this subsurface, produces as output what seismic data would be measured over this surface.  The full-waveform inversion problem is to infer the velocity map from a given set of seismic data. }
\label{fig1: exp}
\end{figure}

\IEEEPARstart{S}eismic full-waveform inversion techniques are commonly used in geophysical exploration to determine site geology, stratigraphy, and rock quality. These techniques provide information about subsurface layering and rock geomechanical properties. In particular, seismic full-waveform inversion infers a 2D/3D map of seismic velocity from observations (Fig.~\ref{fig1: exp}). 
The seismic velocity depends on and therefore predicts subsurface material properties. There are two primary ways of solving this problem, depending on the complexity of the forward model that is used. The simpler approach is via travel time inversion~\cite{tarantola-2005-inverse}, which has a linear forward operator, but provides results of inferior accuracy and resolution~\cite{lin2015Double}. Acoustic- and elastic-waveform inversion (AEWI) techniques~\cite{virieux2009overview, hu2009simultaneous,fichtner2010full,guitton2012constrained} provide superior solutions by modeling the wave propagation in the subsurface, but the forward operator is non-linear and computationally expensive to simulate, and the problem is ill-posed, without a unique solution~\cite{virieux2009overview},

Acoustic- and elastic-waveform inversion~(AEWI) can be solved in either time domain or frequency domain~\cite{vigh2008comparisons, hu2009simultaneous, guitton2012blocky}.   The major challenges of solving AEWI mostly come from three folds:ill-posedness, cycle skipping, and high computational cost. Similar to other geophysical exploration methods, AEWI suffers from the limited data coverage, which results in extremely under-constrained inverse problems. Due to the fact that AEWI is highly non-linear and  sensitive to the initial guess, a naive approach to the AEWI problem  typically converges to a local minima, which is also called cycle skipping.  Having low-frequency components in inversion is critical to alleviate this cycle skipping issue. To make the matter worse. Solving AEWI problems is also computationally expensive. Most of the existing approaches to solve AEWI rely on iterative nonlinear optimization techniques. At each iteration, it costs $\mathcal{O}(n^3)$ to obtain the gradient,  provided with a 2-D $n \times n$ subsurface model.

To mitigate those aforementioned issues, many regularization approaches have been proposed and developed in recent years, which includes Tikhonov-like regularization~\cite{hu2009simultaneous,burstedde2009algorithmic,ramirez2010regularization}, total-variation regularization~\cite{lin-2017-building, Lin-2015-Quantifying, lin2014acoustic,guitton2012blocky,anagaw2011full}, high-order regularization techniques~\cite{treister-2016-full}, and prior-based methods~\cite{ma2012image,zhang2013double}. Most of those existing regularization and prior-based techniques are hand crafted, meaning that are loosely
(if at all) related to the physical problem at hand.  Furthermore, all these solutions are developed under the physics-driven AEWI framework. Hence, the expensive computational costs will be inherited and unavoidable.

More recently, with the successes of deep learning in computer vision community~\cite{zhu2017unpaired, isola2017image, zhang2018boundary}, researchers have developed various data-driven seismic AEWI techniques~\cite{wang2018velocity, Ovcharenko-2018-Low, Richardson-2018-Seismic,Sun-2018-Low, Araya-2018-Deep, wu2018inversionnet}.  Data-driven frameworks take the waveform data as the input and directly outputs its corresponding velocity image. In this work, we study generative adversarial network (GAN)~\cite{goodfellow2014generative} based method. GAN  has been proved to be effective in areas of photo inpainting~\cite{isola2017image, zhu2017unpaired}, image denoising \cite{yang2018low}, super-resolution \cite{ledig2017photo}, image deblurring~\cite{kupyn2017deblurgan}, and so forth. Motivated by these successes, we solve the AEWI problem using GAN. Specifically, our model consists of two parts: generator and discriminator. Generator is an encoder-decoder structure which maps the raw seismic waveform data into velocity image. Discriminator is a convolutional neural network (CNN) designed to classify the real velocity image and fakes velocity image. There are two major benefits using GAN to solve our seismic waveform inversion problems. Firstly, our model learns regularization term directly from data through the discriminator and further impose the learned regularization term to the generator. The regularization term is used to differentiate between ground truth velocity map and generated velocity map. This type of GAN-based regularization has been recently discussed in ~\citet{Li-2018-Learning} and yields supreme results for computer-vision tasks. Secondly, our GAN-based inverse problem model is an end-to-end framework which is similar to image-to-image translation problem from computer vision\cite{gatys2015neural, johnson2016perceptual}, which means the velocity map can be output in real-time once the training is completed. 

Compared with physics-driven methods, the major disadvantage of data-driven methods is that they suffer from robustness and generalization  issue. The deep neural network which is trained on a specific dataset tends to perform worse on another dataset. To alleviate the issue, we incorporate our data-driven method with the network-based deep transfer learning. Network-based deep transfer learning means the reuse of network which is pre-trained in the source domain, and then transfer the network parameters and structures to the target domain. In our project, we apply the fine tuning strategy which means that all of the model's parameters for the new dataset are updated.

To summarize, the main contributions of our work are:
\begin{list}{$\bullet$}
{ \setlength{\leftmargin}{0.18cm}}
	\item To the best of our knowledge, we are the first to apply conditional adversarial network on AEWI. Our model transfers the inverse process of physics-driven methods into an image mapping problem. As a result, it can alleviate the local minima and low computational efficiency issues.
	\item We develop a modified encoder-decoder structure which is more suitable for AWEI. Besides, we combine mean absolute error (mae) loss with mean square error (mse) loss to further improve the quality of velocity images. Compared with the other deep learning baselines, our model is able to generate more accurate velocity images.
	\item We show that the GAN-based regularization technique can yield better reconstruction accuracy than the encoder-decoder-based inversion method. 
    \item We perform a series of experiments to demonstrate the robustness of our model, validating that our model does not just ``memorize'' the training data while it learns the intrinsic physics law from the training set.
    \item We conduct additional experiments to demonstrate that data-driven method plus transfer learning is a feasible way to alleviate the generalization issue.
\end{list}

In the following sections, we first briefly provide the related work in Section~\ref{exp: relatedwork}. We also describe the fundamentals of physics-driven versus data-driven methods, and deep neural networks~(Section~\ref{section: model}). We then develop and discuss our novel inversion method - VelocityGAN. Section~\ref{exp:results} describes the data we tested on, experimental setup, and experimental results we obtained. Finally, concluding remarks are presented in the Conclusions Section.

\section{Related Work}
\label{exp: relatedwork} 

\subsection{Data-driven Inverse Problems}

Acoustic- and elastic-waveform inversion~(AEWI) is a specific type of inverse problems. We first provide relevant literatures in solving inverse problems from other domains. In particular, we focus on deep neural network related techniques\cite{Adler-2018-Learned, Hammernik-2018-Image, Zhu-2018-Image, Coupled-2017-Zeng, Dr2-2017-Yao, Accurate-2015-Kim}. In general, those different deep-learning based methods for solving inverse problems can be categorized into four types~\cite{Deep-2018-Lucas}: 1)~to learn an end-to-end regression with vanilla convolutional neural network~(CNN), 2)~to learn higher-level representation, 3)~to gradual refinement of inversion procedure, and 4)~to incorporate with analytical methods and to learn a denoiser. An interesting work under the first category is AUTOMAP, which was recently developed by \citet{Zhu-2018-Image}. The authors developed an end-to-end reconstruction algorithm for MRI imaging, where the encoder consists of three fully connected network to read in sensor-domain data and the decoder consists of three additional convolutional and de-convolutional layers to yield the image-domain output. A common use of CNNs is to learn a compressed representation prior to constructing an output image. Several existing works use the effectiveness of autoencoders to learn relevant features to solve inverse problems in imaging. As an example, \citet{Coupled-2017-Zeng} employ the autoencoder's representation-learning capability to learn useful representations of low-resolution and high-resolution images. A shallow neural network is then trained to learn a correspondence between the learned low-resolution representation and the high-resolution representation. In the third category, CNNs are used to learn a residual between two or more layers by the skip connection from the input of the residual block to its output. This network structure is particularly well suited to inverse problems such as image restorations when the input and the output images share similar content. The work of \citet{Dr2-2017-Yao} and \citet{Accurate-2015-Kim} both belong to this category. Another type of research effort to solve inverse problems using neural networks is to incorporate analytical solutions. The work developed in \citet{Hammernik-2018-Image} also falls into this category. \citet{Hammernik-2018-Image} reformulates a generalized compressed sensing reconstruction as a variational model, which is embedded in an unrolled gradient descent iterative scheme. Key parameters such as those used in activation functions are learned through offline training procedure. In the inference stage, the previously learned model will be applied online to unseen data. Another example under this category is the one developed in \citet{Adler-2018-Learned}. They unrolled a proximal primal-dual optimization method, and replaced the proximal operators using CNNs, and successfully applied to CT image reconstruction problem. 

\subsection{Data-driven Acoustic- and Elastic-waveform Inversion}

Particularly in seismic waveform inversion, there have some recent development of data-driven waveform inversion techniques, which can be categorized into two groups: an end-to-end learning~\cite{ wang2018velocity, Richardson-2018-Seismic, Araya-2018-Deep, wu2018inversionnet} and low-wave number learning~\cite{Ovcharenko-2018-Low, Sun-2018-Low}. The end-to-end strategy directly learns a mapping correspondence from seismic data domain to the velocity model domain. The low-wave number strategy learns low-wave number from data and followed by traditional full-waveform inversion iteration. Comparing these two strategies, the end-to-end learning strategy is more aggressive, which usually requires much more complex networks structures to account for the nonlinearity nature of the full-waveform inversion. Encouraging results have been recently reported in \citet{wu2018inversionnet} due to significant amount of training sets are utilized.

\subsection{Deep Transfer Learning}
A great number of deep transfer learning methods are developed in the recent years. There are mainly four types of deep transfer learning approaches, which are instance-based deep transfer learning, mapping-based deep transfer learning, network-based deep transfer learning, and adversarial-based deep transfer learning~\cite{tan2018survey}. Our work belongs to the network-based deep transfer learning. Two types of network-based deep transfer learning are widely used in practical applications: fine tuning and feature extraction. Feature extraction refers to the reuse of a pre-trained model and only update a few layer weights for the target domain. For example, the authors in ~\cite{oquab2014learning} reuse front-layer trained on ImageNet to compute intermediate image representation for images in other datasets. Fine tuning means all of the model parameters for a new task are updated.

\section{The Inversion Models}
\label{section: model}
We firstly present some overview of  the governing physics equation~(acoustic and elastic wave equation),  physics-driven AEWI method, and data-driven inversion method in Section \ref{subsec2: AEWI}. In Section \ref{subsec3: structure}, we provide details on our VelocityGAN and its network structure. In Section \ref{subsec4: loss function}, we provide the loss functions of our VelocityGAN. In Section~\ref{regularization}, we discuss the connections to inverse and regularization theory. 

\subsection{Acoustic- and Elastic-Waveform Inversion: Physics-Driven Approach}
\label{subsec2: AEWI}

\subsubsection{Governing Physics - Wave Equation}

Mathematically, the seismic acoustic-wave equation (or ``forward model'') is
\begin{equation}
\left [ \frac{1}{K(\mathbf{r})} \frac{\partial ^2}{\partial t ^2} 
- \nabla  \cdot \left ( \frac{1}{\rho (\mathbf{r})}\,\, \nabla \right 
) \right ]
p(\mathbf{r}, t) = s(\mathbf{r},\, t),
\label{eq:Forward}
\end{equation}
where $\rho (\mathbf{r})$ is the density at spatial location $\mathbf{r}$, $K(\mathbf{r})$ is the bulk modulus, $s(\mathbf{r},\, t)$ is the source term, $p(\mathbf{r}, t)$ is the pressure wavefield, and $t$ represents time. 
The elastic-wave equation is written as
\begin{equation}
\rho(\mathbf{r})\, \ddot{u}(\mathbf{r}, t) - \nabla \cdot [C(\mathbf{r}) 
: \nabla u(\mathbf{r}, t)] = s(\mathbf{r},\, t),
\label{eq:ForwardElastic}
\end{equation}
where $C(\mathbf{r})$ is the elastic tensor, and $u(\mathbf{r}, t)$ is the displacement wavefield.
To simplify the expression, the forward modeling problems in Eqs.~\eqref{eq:Forward} and \eqref{eq:ForwardElastic} can be written as
\begin{equation}
P = f(\mathbf{m}),
\label{eq:ForwardLinearM}
\end{equation}
where  $P$ is the pressure wavefield for the acoustic case or the displacement wavefields for the elastic case, $f$ is the forward acoustic or elastic-wave modeling operator, and $\mathbf{m}$ is the velocity model parameter vector, including the density and compressional- and shear-wave velocities.  We use a time-domain stagger-grid finite-difference scheme to solve the acoustic- or elastic-wave equation. 
Inference of unknown subsurface properties relies on indirect and limited geophysical measure-ments taken at or near the surface. Therefore, seismic full-waveform inversion is extremely under-constrained and can be severely ill-posed. Various explicit regularization techniques have been developed to stabi-lize the computation of seismic inversion. This regularized physics-driven seismic inversion can be posed as 
\begin{equation}
E(\mathbf{m}) = \underset{\mathbf{m}}{\operatorname{min}} \left 
\{\left \| \mathbf{d} - f(\mathbf{m})\right \| _2 ^2 + \lambda\, R(\mathbf{m}) 
\right \},
\label{eq:MisFit}
\end{equation}
where $\mathbf{d}$ represents a recorded/field waveform dataset, 
$f(\mathbf{m})$ is the corresponding forward modeling result, $ \left \| \mathbf{d} - f(\mathbf{m})\right \| _2 ^2$ is the data misfit,  
$||\cdot ||_2$ stands for the $\text{L}_2$ norm, $\lambda$ is a regularization parameter and $R(\mathbf{m})$ is 
the regularization term. The regularization term measures the ``complexity'' of the model $f(\mathbf{m})$ so that the minimization in Eq.~\eqref{eq:MisFit} favors simple solution that are consistent with the data. Explicit regularization techniques such $\text{L}_1$-norm~\cite{anagaw2011full, guitton2012blocky, lin2014acoustic, Lin-2015-Quantifying} or $\text{L}_2$-norm~\cite{ramirez2010regularization, fichtner2010full, Treister-2017-Full} have been developed for seismic inversion, but these regularizers do not fully express an expert's prior knowledge.

\subsubsection{Data-driven Approach}

Different from the physics-driven methods, end-to-end data-driven methods transfer the minimization process into a mapping problem. The parameters of $\mathbf{m}$ are directly learned from
\begin{equation}
    \mathbf{m} = g(\mathbf{d}) = f^{-1}(\mathbf{d}),
\end{equation}
where $g = f^{-1}(\cdot)$ is the inverse operator of $f(\cdot)$. We can therefore obtain the loss function as below
\begin{equation}
g = \argmin_{g}\left \{ \sum_{i=1} ^{N} \|\mathbf{m}_{i}  -  g(\mathbf{d}_{i})  \|_2^2 \right \},
\label{eq:encoder-decoder}
\end{equation}
where $(\mathbf{m}_{i}, \mathbf{d}_{i})$ are $N$ pairs of velocity map and the corresponding seismic data. Most of the existing end-to-end data-driven AEWI methods use encoder-encoder structure to learn the mapping function of $g$ in Eq.~\eqref{eq:encoder-decoder}~\cite{wang2018velocity, Richardson-2018-Seismic, Araya-2018-Deep, wu2018inversionnet}.

\subsection{VelocityGAN: Data-Driven Approach}
\label{subsec3: structure}

\begin{figure*}[t]
\begin{center}
    \includegraphics[width=0.6\linewidth]{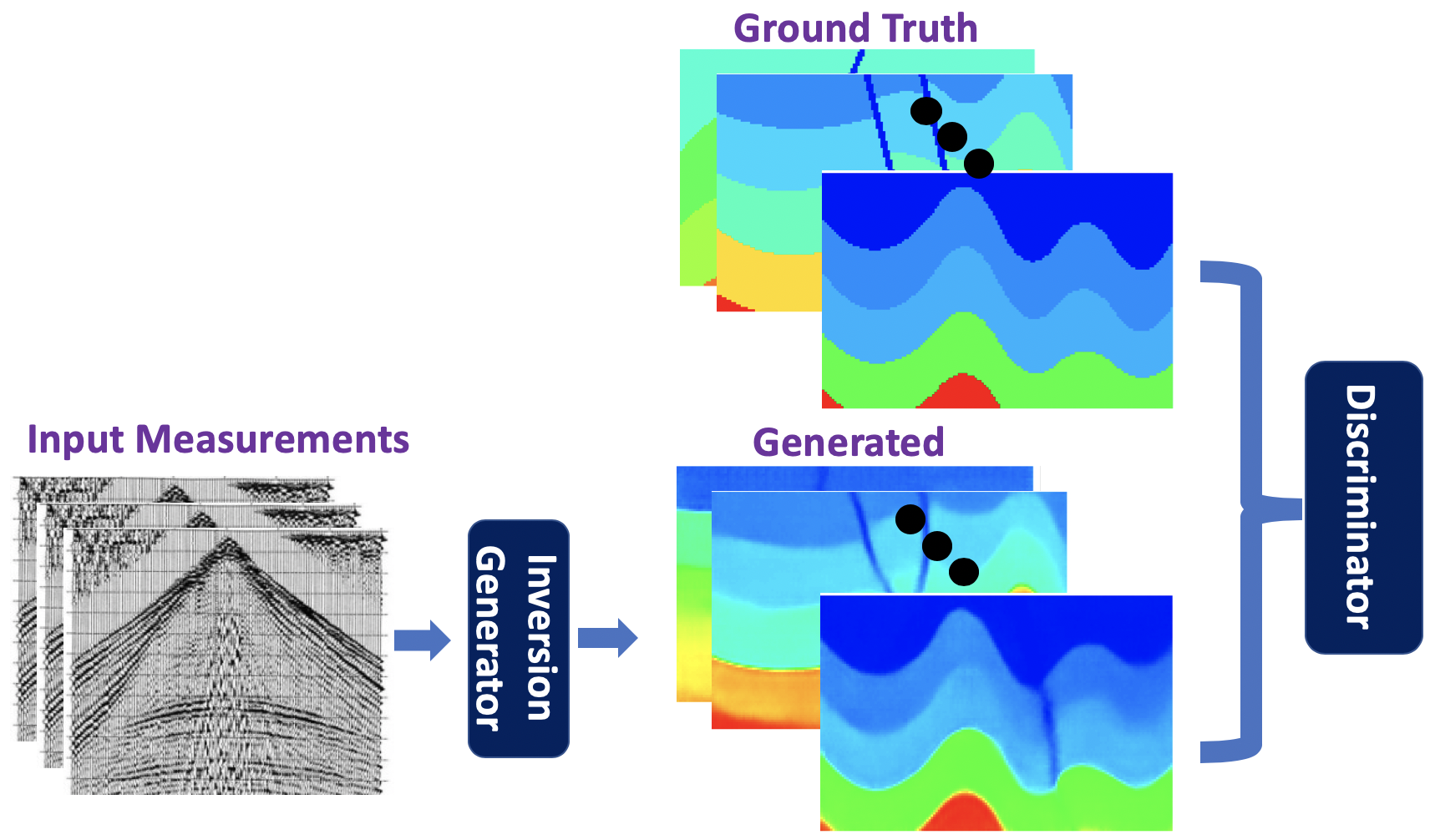}
\end{center}
   \caption{The overall architecture of VelocityGAN. We apply an encoder-decoder structure as the generator (``Inversion Generator'') and a convolutional neural network as the discriminator.}
\label{fig2: whole structure}
\end{figure*}

The overall architecture of VelocityGAN is shown in Fig.~\ref{fig2: whole structure}. It consists of an image-to-image generator and a CNN discriminator. We discuss below the structure of the generator and the discriminator, respectively. 

\subsubsection{Generator}
\begin{figure}[h]
\begin{center}
    \includegraphics[width=1.0\linewidth]{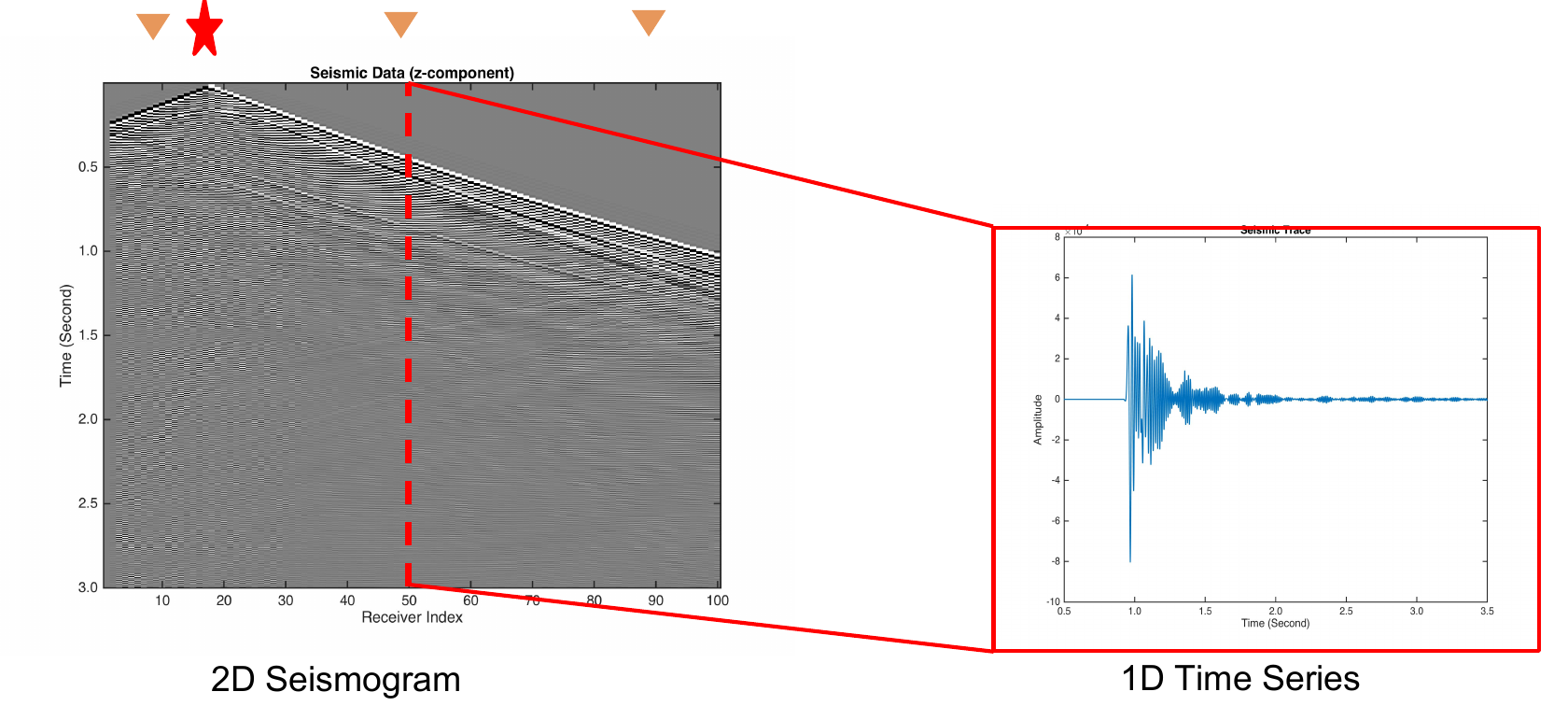}
\end{center}
   \caption{The visualization of 2D seismogram. The value in horizontal direction represents different receivers. The value in vertical direction represents the 1D time series signal. We pick out a 1D time series signal and present it in the red bounding box.}
\label{fig3: data visualization}
\end{figure}

\begin{figure*}[t]
\begin{center}
    \includegraphics[width=1.0\linewidth]{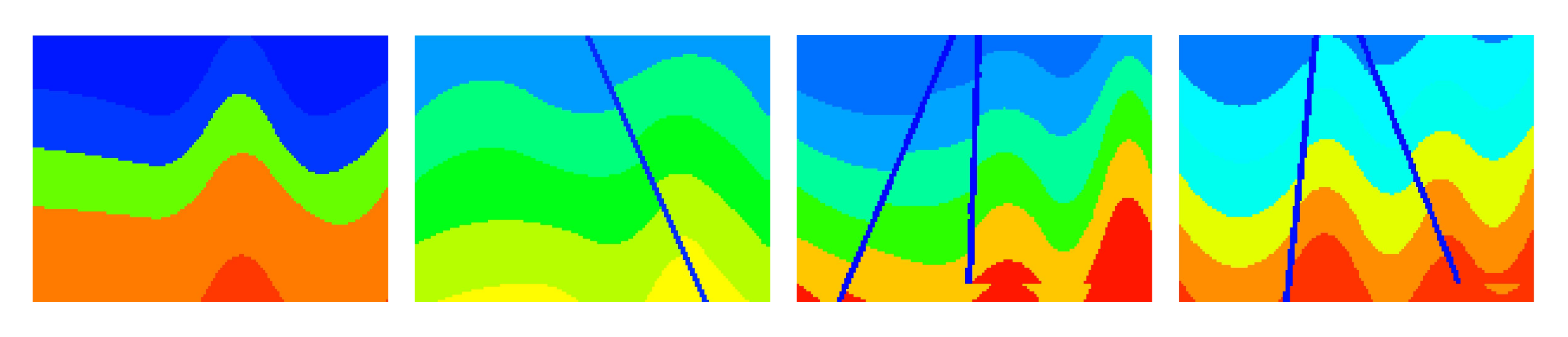}
\end{center}
   \caption{The visualization of velocity images. The number of geologic faults (the linear features shown in the velocity map) varies with different subsurface structure.}
\label{fig4: data visualization}
\end{figure*}

To better understand the network structure, we first recall the governing physics of our input seismic shot-gather imagery~(Fig.~\ref{fig3: data visualization}) and output velocity map~(Fig.~\ref{fig4: data visualization}). In this paper, the input seismic shot-gather imagery is the combination of the acoustic waves received by different receivers and the visualization of an exemplar 2D seismogram is shown in Fig.~\ref{fig3: data visualization}.  Specifically, there are 3 source functions and 32 receivers, which correspond to $s(\mathbf{r}, t)$ and $P(\mathbf{r}, t)$ in Eq.~\eqref{eq:Forward} (or $u(\mathbf{r}, t)$ in Eq.~\eqref{eq:ForwardElastic}), respectively. The source function may contain both P-wave and S-wave. Correspondingly, the seismic waveform data collected is a tensor with a dimension of $32 \times 1000 \times 6$, where the first dimension is 32 receivers, the second dimension of 1000 is the time sequence length of waveform trace, and the third dimension of 6 represents the total 2 channels of the 3 source functions. As the shown in Fig.~\ref{fig3: data visualization}, the 2D seismogram consists of 32 1D time series signals, and each signal contains a pulse which contains the information of subsurface structure. Correspondingly, we also present several velocity images in Fig~\ref{fig4: data visualization}, which is the output of our model. The dimension of the output velocity map is $(m, n)$, where the first dimension of $m$ stands for depth and the second dimension of $n$ stands for horizontal offset. The value of each pixel in the velocity image stands for the absolute velocity value at each location. The grid spacing between pixels is 5 meters. Therefore, the total size of velocity map in the real world is $5m\times5n$ meters. The linear geologic feature shown in the velocity map in Fig.~\ref{fig4: data visualization} is the geologic fault.

As we discussed above, there is no direct spatial similarities between the input seismic gather-shot imagery and output subsurface velocity map. See Figs.~\ref{fig3: data visualization} and \ref{fig4: data visualization}. So we do not penalize the mismatch between the input and output like \cite{ronneberger2015u, zhu2017unpaired}. Besides, since the height (1000) and width (32) of the input is unbalanced, we apply several convolutional layers with $k\times1$ kernels, $k$ means the length of convolutional kernel in height dimension. The particular structure of generator is shown in Table \ref{table1: generator architecture}. To extract the waveform features of each receiver, 9 convolutional layers with $k\times1$ kernels are first deployed. Each convolutional layer is followed by a BatchNormalization layer and a LeakyReLU layer. After the dimension of height is reduced to 32, $3\times3$ convolutional kernels with stride 2 are then added to encode the whole extracted features. In the last layer of encoder, $8\times8$ convolutional kernels are used to eliminate the influence of spatial information. With regards to the decoder, it consists of 5 upconv blocks, a center cropping layer, and a convolutional layer. Each upconv block consists a transposed convolutional layer, batch normalization layer, and an activation layer. The transposed conlutional layers are applied to increase the height and width dimensions of image and decode the extracted features. The convolutional layer is designed to map features into the same dimension with ground truth labels. The center cropping layer is used to crop the feature maps into a desired dimension. To limit the value of output into a specific range, the center cropping layer is followed by a Tanh layer. We have also tried to replace the transposed convolution with upsampling layer like \citet{ronneberger2015u} and \citet{shelhamer2016fully}. It yields worse performance because the geologic fault, i.e., the linear geologic feature,  of velocity map is not as good as these structure.

\begin{table}
\renewcommand\arraystretch{1.5}
\caption{The generator architectures of VelocityGAN.}.
\label{table1: generator architecture}
\begin{tabu}{c|c|c}
\hline
Layers & Output Size & The Generator of VelocityGAN \\
\hline
conv1 & 500$\times$32 & 7$\times$1 conv, channel 32, stride 2$\times$1 \\
\hline
conv block1 & 125$\times$32 & 
$\left[ \begin{array}{c} 3\times1, 64, 2\times1  \\ 3\times1, 64, 1\times1 \end{array} \right] \times2 $ \\
\hline 
conv block2 & 32$\times$32 & 
$\left[ \begin{array}{c} 3\times1, 128, 2\times1  \\ 3\times1, 128, 1\times1 \end{array} \right] \times2 $ \\ 
\hline
conv block3 & 8$\times$8 & 
$ \left[ \begin{array}{c} 3\times3, 256, 2\times2  \\ 3\times3, 256, 1\times1 \end{array} \right] \times2 $ \\ 
\hline
conv2 & 1$\times$1 & 
8$\times$8, 512, 1$\times$1 \\
\hline
upconv block1 & 7$\times$7 & 
$\left[ \begin{array}{c}  7\times 7\quad{\rm deconv}, 512, 2\times2  \\ 3\times3\quad{\rm conv}, 512, 1\times1 \end{array} \right] $  \\
\hline
upconv block2 & 14$\times$14 & 
$\left[ \begin{array}{c} 4 \times 4  , 256, 2\times2  \\ 3\times3, 256, 1\times1 \end{array} \right] \times2 $ \\
\hline
upconv block3 & 28$\times$28 & 
$\left[ \begin{array}{c} 4 \times 4  , 128, 2\times2  \\ 3\times3, 128, 1\times1 \end{array} \right] \times2 $ \\
\hline
upconv block4 & 56$\times$56 & 
$\left[ \begin{array}{c} 4 \times 4  , 64, 2\times2  \\ 3\times3, 64, 1\times1 \end{array} \right] \times2 $ \\
\hline
upconv block5 & 128$\times$128 & 
$\left[ \begin{array}{c} 4 \times 4  , 32, 2\times2  \\ 3\times3, 32, 1\times1 \end{array} \right] \times2 $ \\
\hline
conv3 & 100$\times$100 & 3$\times$3, 1, 1$\times$1 \\
\hline
crop layer & 100$\times$100 & --- \\
\hline
\end{tabu}
\end{table}

\subsubsection{Discriminator}
Similar to \citet{radford2015unsupervised}, we adapt our discriminator from a CNN architecture. Particularly, it consists of five convolution blocks, a global average pooling layer, and fully connected layers. Each convolutional block involves a combination of Convolutional, BatchNormalization, LeakyReLU, and MaxPooling layer. We apply ``PatchGAN'' classifer~\cite{isola2017image} in the discriminator to capture local style statistics. We set the patch size as 4 and calculate the mean loss value of all patches in an image. Since we would like to generate accurate velocity models, especially in the part of geological fault and interfaces, ``PatchGAN'' is more suitable than ``GlobalGAN'' for our task.

\subsection{Loss Function}
\label{subsec4: loss function}
Wasserstein GAN (WGAN) with gradient penalty \cite{gulrajani2017improved} has been proved to be robust of a wide variety of generator architectures. Considering the modified structure in our generator, we use Wasserstein loss with gradient penalty to distinguish the real velocity map and generated velocity map. The loss function of discriminator is formulated as
\begin{equation}
    L_d = \underset{\Tilde{x} \sim \mathbb{P}_g}{\mathbb{E}} D(\Tilde{x}) - \underset{x\sim\mathbb{P}_r}{\mathbb{E}}D(x) + \lambda\underset{\hat{x}\sim\mathbb{P}_{\hat{x}}}{\mathbb{E}}[(\| \nabla_{\hat{x}}D(\hat{x})\|_2-1)^2],
    \label{eq: disminator loss}
\end{equation}
where $\mathbb{P}_g$ means the distribution of velocity map which is predicted by the generator of VelocityGAN, $\mathbb{P}_r$ is the distribution of the ground truth velocity map, and $\mathbb{P}_{\hat{x}}$ is random samples from both $\mathbb{P}_g$  and $\mathbb{P}_r$.

For the generator, we want the predicted velocity map can not only fool the discriminator but also reveal the accurate information of geological structure. Therefore, the loss function is a combination of the adversarial loss and content loss. Consistent with Eq. \eqref{eq: disminator loss}, the adversarial loss is $-\underset{\Tilde{x}\sim\mathbb{P}_g}{\mathbb{E}}D(\Tilde{x})$. The content loss is set as a combination of mean absolute error (mae) and mean square error (mse). In our experiments, we observe that mse loss is good at capturing the geological faults while mae loss performs better on revealing the geological interfaces. Therefore, the loss function of generator is formulated as
\begin{equation}
\begin{split}
    L_g = & -\underset{\Tilde{x} \sim \mathbb{P}_g}{\mathbb{E}} D(\Tilde{x}) + \frac{\lambda_1}{w\cdot h}\sum_{i=1}^{w}\sum_{j=1}^{h}|\Tilde{v}(i,j)-v(i,j)| \\ 
     &+ \frac{\lambda_2}{w\cdot h}\sum_{i=1}^{w}\sum_{j=1}^{h}(\Tilde{v}(i,j)-v(i,j))^2,
    \label{eq: generator loss}
\end{split}
\end{equation}
where $w$ and $h$ are the width and height of the velocity map respectively, $v(\cdot)$ represents the real pixel value of the velocity map and $\Tilde{v}(\cdot)$ means the predicted pixel value. $\lambda_1$ and $\lambda_2$ are hyper-parameters to control the relative importance of the two loss term. In our experiments, we pick $\lambda_1$ and $\lambda_2$ by balancing mae and mse loss during the training process. The specific values of $\lambda_1$ and $\lambda_2$ are discussed in Section \ref{Exp: Training Details}.

\subsection{Connection to Regularization Theory}
\label{regularization}
There is a close connection of GAN to the regularization techniques used in inverse problems~\cite{Li-2018-Learning}. To see the connection, we can rewrite Eq.~\eqref{eq: generator loss} as
\begin{eqnarray}
\nonumber
g = \argmin_{g}\{\lambda_1 \sum_{i=1} ^{N} \|\mathbf{m}_{i}  -  g(\mathbf{d}_{i})  \|_1  + \lambda_2 \sum_{i=1} ^{N} \|\mathbf{m}_{i}  -  g(\mathbf{d}_{i})  \|_2^2 \\
-\underset{\Tilde{x} \sim \mathbb{P}_g}{\mathbb{E}} D(\Tilde{x}) \},
\label{eq:GAN_loss}
\end{eqnarray}
where the target mapping, g, in Eq.~\eqref{eq: generator loss} can be interpreted as either the inversion operator according to Eq.~\eqref{eq:GAN_loss} or the generator according to Eq.~\eqref{eq: generator loss}. Similarly, the term of $-\underset{\Tilde{x} \sim \mathbb{P}_g}{\mathbb{E}} D(\Tilde{x})$ is not only an adversarial loss term but also plays the role of regularization, that is learned from training data.The content-loss terms~(mae and mse) in Eq.~\eqref{eq: generator loss} or Eq.\eqref{eq:GAN_loss} promotes the velocity-map consistency.  In particular, we will use GANs to learn a classifier to discriminate between the distribution of the ground truth velocity maps and the distribution of generated velocity maps. This discriminator effectively penalizes velocity models that do not ``look like'' the velocity models that are used for training. The usual approach to alleviate ill-posedness of inverse problems is to incorporate prior knowledge with a regularization term that penalizes solutions that are inconsistent with this prior knowledge. Most of the existing regularization techniques employ generic functions (e.g., $\text{L}_1$-norm or $\text{L}_2$-norm penalties on coefficients) that are loosely (if at all) related to the physical problem at hand. On the other hand,  regularization learned from data can be more effective and customized for the problem at hand.

\section{Experiments}
\label{exp:results}

We introduce the datasets and training details in Section \ref{exp: datasets}. We discuss the experiment settings in Section \ref{exp: expsettings}. Following that, we compare and analyze the results of different methods. Last but not least, we present generalization experiments and provide a feasible way to solve the generalization issue.

\subsection{Datasets and Training Details}
\label{exp: datasets}
\subsubsection{Datasets}

In practical applications, velocity models are estimated by physics-driven methods (usually an optimization algorithm). It can be unrealistic and expensive to obtain a large-scale dataset consists of seismic waveform and velocity models. To verify the efficacy of our VelocityGAN, we therefore generate a dataset including velocity images and corresponding seismic waveform data generated using Eq.~\eqref{eq:Forward}.  The velocity images that we generated are varied with different tilting angles, layer thicknesses, and layer velocities etc. They can be a good representation of the real velocity images~\cite{Lin2018Efficient}. Although our VelocityGAN is validated using seismic acoustic wave equation, the method developed in this paper can be directly adapted to elastic scenario shown in Eq.~\eqref{eq:ForwardElastic} as well. 

We create a main dataset to evaluate the efficiency and effectiveness of VelocityGAN. The dataset contains 50,000 velocity models with 150 by 100 dimension, along with their corresponding seismic waveform. This dataset contains complicated geological layers in a velocity image. Furthermore, most geological layer interfaces are curved. We name this dataset as ``CurvedData''. For this dataset, 3 common-shot gather of synthetic seismic data with $32$ receivers is posed at the top surface. We use a Ricker wavelet with a center frequency of $50$~Hz as the source time function and a staggered-grid finite-difference scheme with a perfectly matched layered absorbing boundary condition to generate 2D synthetic seismic reflection data~\cite{Tan-2014-Efficient, Zhang-2010-Unsplit}. The synthetic trace at each receiver is a collection of time-series data of length $1,000$. Hence, the input size is (32, 1000, 6) where 1000 is the time sequence length, 32 is the number of receivers and 6 is the channel number. 

We also create two small datasets for generalization experiments. The velocity images of these two datasets are similar with the images in CurvedData except for the number of faults. We adjust the number of geological faults to zero or two in order to evaluate our VelocityGAN in a more generalized condition. We use 2-Fault CurvedData and 0-Fault CurvedData to represent these two datasets in the following parts. There are 2,000 pairs of velocity images in 2-Fault CurvedData and 1,000 pairs of velocity images in 0-Fault CurvedData. 

\subsubsection{Training Details}
\label{Exp: Training Details}
For each dataset, we randomly select 20\% data as the testing set, 10\% data as the validation set to adjust the hyperparameters. We use the remaining images as the training set. The input of our model is normalized to range $(-1\sim1)$. Constrained by the memory of GPU, we set the size of mini-batch to 50. Following the optimization strategy of \cite{arjovsky2017wasserstein}, we perform 5 gradient descent steps on the discriminator, and then perform one step on the generator. The learning rate of our Adam~\cite{kingma2014adam} optimizer is set to $10^{-4}$ in the first epoch. We linearly decay the learning rate to 0 over the remaining epochs. For discriminator loss (Eq.~\eqref{eq: disminator loss}), we choose $\lambda$ as 10. For generator loss (Eq.~\eqref{eq: generator loss}), $\lambda_1$ and $\lambda_2$ are set to  50 and 100 in CurvedData. All of our models are implemented on a single GTX 1080-Ti using PyTorch framework.

\begin{table*}
\renewcommand\arraystretch{1.5}
\caption{ Quantative Results of Velocity Image Reconstruction on CurvedData}.
\label{table3: Quantitative-CurvedData}
\centering
\begin{tabular}{c|ccc|cccc}
 & mae & rel ($10^{-3}$) & log10 ($10^{-3}$) & acc. (t=1.01) &acc. (t=1.02) &acc. (t=1.05) &acc. (t=1.10) \\
\hline
AEWI-Pre \cite{zhang2012wave} & 144.87 & 60.28 & 26.33 & 24.05\% & 38.09\% & 62.66\% & 81.38\% \\
AEWI-MTV \cite{lin2014acoustic} & 136.80 & 56.25 & 24.44 & 35.56\% & 47.30\% & 67.69\% & 81.20\%\\
\hline
Generator-l1 & 83.03 & 36.77 & 14.97 & 40.05\% & 59.16\% & 84.59\% & 93.41\% \\ 
Generator-l2 & 80.93 & 34.43 & 14.64 & 34.29\% & 56.64\% & 84.69\% & 94.16\% \\ 
Generator & 85.69 & 35.99 & 15.44 & 35.99\% & 56.14\% & 82.80\% & 93.35\% \\
\hline
VelocityGAN-l1 & 79.08 & 34.76 & 14.46 & \textbf{42.30}\% & \textbf{62.70}\% & 85.70\% & 93.26\% \\
VelocityGAN-l2 & 76.19 & \textbf{32.39} & 13.91 & 32.39\% & 59.60\% & 86.17\% & \textbf{94.46}\% \\
VelocityGAN & \textbf{75.85} & 32.40 & \textbf{13.83} & 40.64\% & 62.54\% & \textbf{86.33}\% & 93.98\% \\
\end{tabular}
\end{table*}

\subsection{Experiment Settings}
\label{exp: expsettings}

Velocity image generation experiments are conducted to evaluate the effectiveness of  our VelocityGAN. We choose the following algorithms from both physics-driven and data-driven methods as our baselines:

\begin{list}{$\bullet$}
{ \setlength{\leftmargin}{0.4cm}}
    \item AEWI-Pre~\cite{zhang2012wave}: A wave-energy-based precondition method is applied to reduce the artifacts in the gradients caused by the geometrical spreading and defocusing effects.
    \item AEWI-MTV~\cite{lin2014acoustic}:  A modified total-variation regularization (MTV) is used as a regularization term in AEWI optimization process. MTV is designed to preserve sharp interfaces in piecewise constant structures.
	\item U-Net~\cite{ronneberger2015u}: Based on a typical encoder-decoder structure, U-Net adds skip connections between mirrored layers in the encoder and decoder stacks.
	\item FCN~\cite{shelhamer2016fully}: FCN consists of an encoder network and a corresponding decoder network. We apply upsampling layers in the decoder network and adjust the dimension of convolutional layers to make it work on our project.
	\item modifiedFCN~\cite{wang2018velocity}: A modified FCN framework for full waveform inversion (FWI) problem.
\end{list}

Specifically, AEWI-Pre~\cite{zhang2012wave} and AEWI-MTV~\cite{lin2014acoustic} are physics-driven methods. FCN~\cite{shelhamer2016fully}, U-Net~\cite{ronneberger2015u} and modifiedFCN~\cite{wang2018velocity} are selected as the data-driven baselines for solving AEWI problem.

Similar with the existing works~\cite{eigen2015predicting}, \cite{eigen2014depth} on depth estimation, we adopt the following metrics to evaluate the accuracy of velocity image reconstruction:
\begin{list}{$\bullet$}
{ \setlength{\leftmargin}{0.4cm}}
    \item mean absolute error (mae):
        $\mathrm{mae} = \frac{1}{n}\sum_i|m_i-m_i^*|$,
    \item mean relative error (rel):
    $\mathrm{rel} = \frac{1}{n}\sum_i\frac{|m_i-m_i^*|}{m_i^*}$,
	\item mean $\log$10 error($\log$10): 
	$\mathrm{log10} = \frac{1}{n}|\log_{10}m_i-\log_{10}m_i^*|$,
	\item the percentage of $m_i$ (acc.): $\mathrm{acc}  = $max$(\frac{m_i^*}{m_i},\frac{m_i}{m_i^*})<t$.
\end{list}

For qualitative experiments, we present several velocity image samples and vertical velocity profiles to provide an intuitive comparison. It is worthwhile to mention that in  CurvedData, we include a small geologic fault in the velocity model. Geologic faults play an important role in siting the wells in subsurface applications because of its high permeability property. However, it can be technical challenging to image a geologic fault zone due to the limited imaging resolution and data coverage. We will compare our method to others not only in the overall reconstruction quality, but also in the local region such as fault zone. Besides, we also compare the implementation time between physics-driven methods and data-driven methods.

\subsection{CurvedData}
In the real world, the geological layers are usually yields irregular shape. To address the curved layer estimation, we create a challenging dataset --- CurvedData. Using this data, geological faults will disappear with the constraint of mae loss. Mse loss is good at revealing geological faults but does not perform well on reconstructing the layer interfaces. Therefore, we use a combination of mae and mse loss to generate more accurate velocity images. In our quantitative experiments, we not only compare VelocityGAN with physics-driven models but also do ablation study on the combination of loss.

\subsubsection{Quantitative Results}
Table \ref{table3: Quantitative-CurvedData} shows the quantitative results of our ablation study on CurvedData. We can see that our proposed models still perform much better than the physics-driven models. Among our proposed models, we conduct the ablation study. We observe that VelocityGAN with a combination of mae and mse loss can get better prediction results than a single loss. Though the VelocityGAN with a single loss achieve relatively higher scores in some measurements such as rel, acc. (t=1.10), the VelocityGAN with a combination of mae and mse loss obtains a better trade-off under all the metrices. The quantitative experiments on  CurvedData validate that the generator structure, adversarial training strategy, and the combination of losses are all useful improvements and can boost the reconstructed accuracy.

\subsubsection{Qualitative Results}
\begin{figure*}[ht]
\begin{center}
    \includegraphics[width=1.0\linewidth]{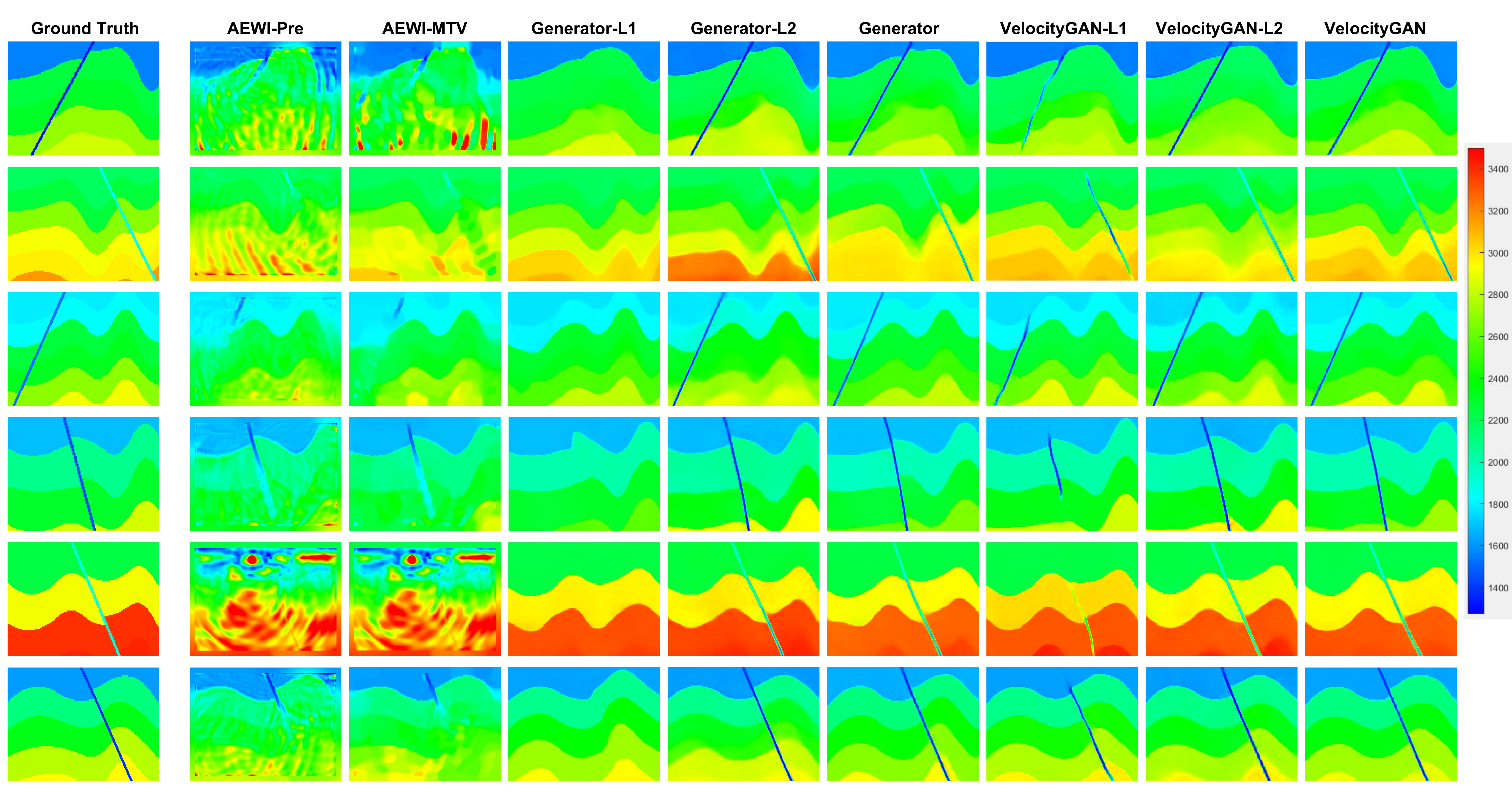}
\end{center}
   \caption{Examples of different methods on CurvedData. The images from left most columns to right are ground truth, reconstruction results using AEWI-Pre~\cite{zhang2012wave}, AEWI-MTV~\cite{lin2014acoustic}, Generator-L1, Generator-L2, Generator, VelocityGAN+L1, VelocityGAN+L2, and VelocityGAN. Our VelocityGAN yields the most accurate reconstructed velocity images among both the physics-driven methods and data-driven baselines. The experiment results substantiate the effectiveness of the adversarial training strategy and a combined loss. }
\label{fig7: CurvedData_eg}
\end{figure*}

We illustrate reconstructions of velocity images using different baseline methods in Figure \ref{fig7: CurvedData_eg}. Consistent with our discussion on loss function, VelocityGAN with mae loss is good at handling with the boundary of geological layers, however, it ignores the geological faults and high-velocity areas. VelocityGAN with mse loss can generate geological faults but the boundary of geological layers are fuzzy. VelocityGAN with a combination of mae and mse can achieve a better tradeoff between the quality of geological layer interfaces and faults. Besides, physics-driven methods does not perform well on CurvedData. There are many oscillations in deep region and high-velocity areas. An interesting find in our experiments is that the comparison of Generator with mae (L1) loss and VelocityGAN with mae loss. For Generator with mae loss, the geological fault disappears in the velocity images. However, VelocityGAN with mae loss can still reveal correct geological fault, though it is fuzzy in some particular area. This phenomenon further demonstrate the effectiveness of adversarial training strategy.

\begin{figure*}[ht]
\begin{center}
    \includegraphics[width=1.0\linewidth]{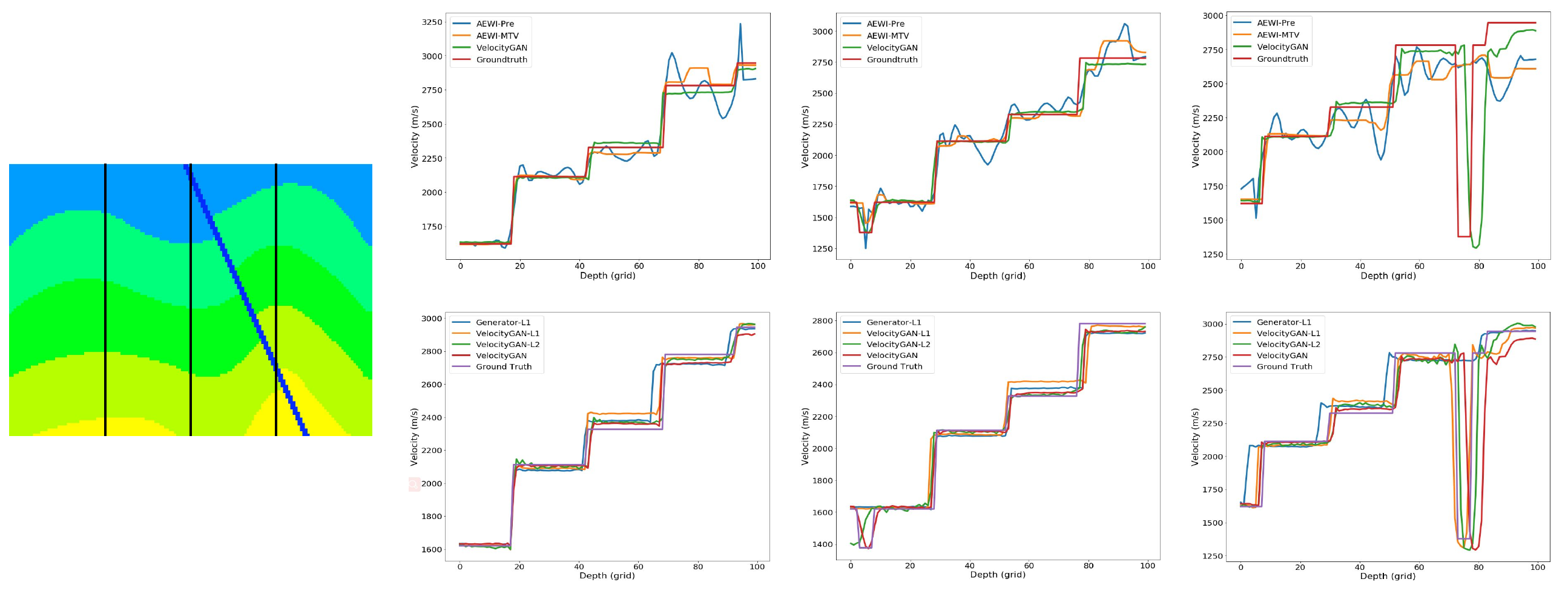}
\end{center}
   \caption{Vertical velocity profiles of different positions on CurvedData. From left to right, the positions are 40, 75, 110 respectively. We compare VelocityGAN with physics-driven methods in the first row, compare VelocityGAN with different loss functions in the third row.}
\label{fig8: vertical_velocity_profile_curve}
\end{figure*}

We present the vertical velocity profile of CurvedData in Fig. \ref{fig8: vertical_velocity_profile_curve}. VelocityGAN still outperforms physics-driven methods on the accuracy of vertical velocity. For VelocityGAN with different loss functions, we observe that the framework with a combination of mae and mse perform well in both low-velocity regions and high-velocity regions. VelocityGAN with a single loss sometimes miss the geological fault zones in low-velocity regions. For example, in the second row and the middle column of Fig.\ref{fig8: vertical_velocity_profile_curve}, both VelocityGAN-L1 and VelocityGAN-L2 fail to reconstruct the geological fault zone between position 0 and 10. In contrast, VelocityGAN with a combination L1(mae) and L2(mse) loss reveal the geological fault well. In the second row of Fig.\ref{fig8: vertical_velocity_profile_curve}, we also compare VelocityGAN-L1 with Generator-L1. Though both of them do not perform well in the low-velocity regions, VelocityGAN-L1 is able to reveal geological fault in high-velocity regions, which is better than Generator-L1. We attribute this phenomenon as the influence of adversarial training strategy. 

According to the aforementioned comparison, we conclude that our VelocityGAN yields more accurate reconstruction of velocity image in obtaining both global and location geological features.

\subsection{Generalization Experiments}

In machine learning theory, the test error of a trained model on unseen data is given as~\citep{Hastie-2016-Elements} 
\begin{equation}
Error_{test} = Error_{train} + Error_{Generalization},
\end{equation}
where $Error_{train}$ is the training error and $Error_{Generalization}$ is the generalization error. With large amount of training data and reasonable loss function, we can usually control the training error, while the generalization error will then dominate the test error. It is well known that deep neural networks are over parameterized meaning there is significantly larger  number of parameters than the amount of training data. Minimizing the same loss function might lead to multiple global minima, which all minimize the training error, but some of them might not generalize well. Conventionally in machine learning community, cross validation techniques are usually used to measure the test error. However, in our problem, cross validation may be misleading  due to the fact that no matter how to split the data, all the training, validation, and test data come from the same distribution. Therefore, we analyze the generalizability of VelocityGAN by studying its performance using specially designed test sets, which are inspired by actual field experiments. 

In CurvedData, all velocity images contain one fault. To conduct the generalization experiments, we generate extra velocity models and their corresponding seismic data as our transfer learning data. Specifically, the transfer learning data includes 0-Fault CurvedData and 2-Fault CurvedData. Based on these two datasets, we compare the reconstruction results of physics-driven methods, VelocityGAN which is trained on CurvedData (VelocityGAN-org), VelocityGAN with transfer learning (VelocityGAN-TF), and VelocityGAN without transfer learning (VelocityGAN). The specific procedure of the transfer learning strategy is that we first train VelocityGAN on CurvedData and save the model weights. We then finetune the VelocityGAN weights on 0-Fault CurvedData or 2-Fault CurvedData.

The quantitative results of generalization experiment are presented in Table \ref{table4: Quantitative-0FaultCurveData} and Table \ref{table5: Quantitative-2FaultCurveData}. In Table \ref{table4: Quantitative-0FaultCurveData}, VelocityGAN-TF outperforms AEWI-Pre, VelocityGAN, and VelocityGAN-org. AEWI-MTV gets a better results than VelocityGAN-TF in 0-Fault CurvedData. For 2-Fault CurvedData, VelocityGAN-TF achieves the best quantitative results of all these models. We attribute the reason as the size of training dataset. For 0-Fault CurvedData, there are 700 pairs of velocity images and seismic data as the training set. In contrast, 2-Fault CurvedData has 1,400 pairs of velocity images and seismic data as the training set, which is two times larger than 0-Fault CurvedData. According to the experiments, we can conclude that the performance of data-driven methods depend on the size of training set. A bigger dataset can achieve a better results. Overall, in our dataset, the performance of VelocityGAN-TF is competitive with the AEWI-MTV while outperforming all others.

\begin{table*}[ht]
\renewcommand\arraystretch{1.5}
\caption{ Quantative Results of Velocity Image Reconstruction on 0-Fault CurvedData}.
\label{table4: Quantitative-0FaultCurveData}
\centering
\begin{tabular}{c|ccc|cccc}
 & mae & rel ($10^{-3}$) & log10 ($10^{-3}$) & acc. (t=1.01) &acc. (t=1.02) &acc. (t=1.05) &acc. (t=1.10) \\
\hline
AEWI-Pre \cite{zhang2012wave} & 74.40 & 29.26 & 13.06 & 39.97\% & 59.68\% & 82.97\% & 93.33\% \\
AEWI-MTV \cite{lin2014acoustic} & \textbf{49.83} & \textbf{19.09} & \textbf{8.52} & \textbf{59.41}\% & \textbf{73.99}\% & \textbf{88.33}\% & \textbf{96.31}\%\\
\hline
VelocityGAN~\cite{ronneberger2015u} & 154.1138 & 67.19 & 28.78 & 11.10\% & 21.36\% & 47.19\% & 76.15\% \\ 
VelocityGAN-org~\cite{shelhamer2016fully} & 82.94 & 34.06 & 14.47 & 31.14\% & 53.54\% & 80.24\% & 92.90\% \\ 
VelocityGAN-TF~\cite{wang2018velocity} & 62.94 & 25.83 & 11.10 & 40.74\% & 63.00\% & 87.29\% & 95.85\% \\
\end{tabular}
\end{table*}

\begin{table*}[ht]
\renewcommand\arraystretch{1.5}
\caption{ Quantative Results of Velocity Image Reconstruction on 2-Fault CurvedData}.
\label{table5: Quantitative-2FaultCurveData}
\centering
\begin{tabular}{c|ccc|cccc}
 & mae & rel ($10^{-3}$) & log10 ($10^{-3}$) & acc. (t=1.01) &acc. (t=1.02) &acc. (t=1.05) &acc. (t=1.10) \\
\hline
AEWI-Pre \cite{zhang2012wave} & 194.43 & 87.29 & 36.68 & 17.53\% & 28.47\% & 50.11\% & 68.36\% \\
AEWI-MTV \cite{lin2014acoustic} & 147.02 & 63.49 & 27.19 & \textbf{31.66}\% & 45.61\% & 66.01\% & 79.32\%\\
\hline
VelocityGAN~\cite{ronneberger2015u} & 242.33 & 107.07 & 45.14 & 7.90\% & 14.82\% & 32.83\% & 55.33\% \\ 
VelocityGAN-org~\cite{shelhamer2016fully} & 128.14 & 56.26 & 23.57 & 29.58\% & 47.25\% & 71.69\% & 86.39\% \\ 
VelocityGAN-TF~\cite{wang2018velocity} & \textbf{115.87} & \textbf{53.10} & \textbf{21.48} & 30.56\% & \textbf{49.32}\% & \textbf{75.87}\% & \textbf{88.30}\% \\
\end{tabular}
\end{table*}

\begin{figure*}[ht]
\begin{center}
    \includegraphics[width=0.95\linewidth]{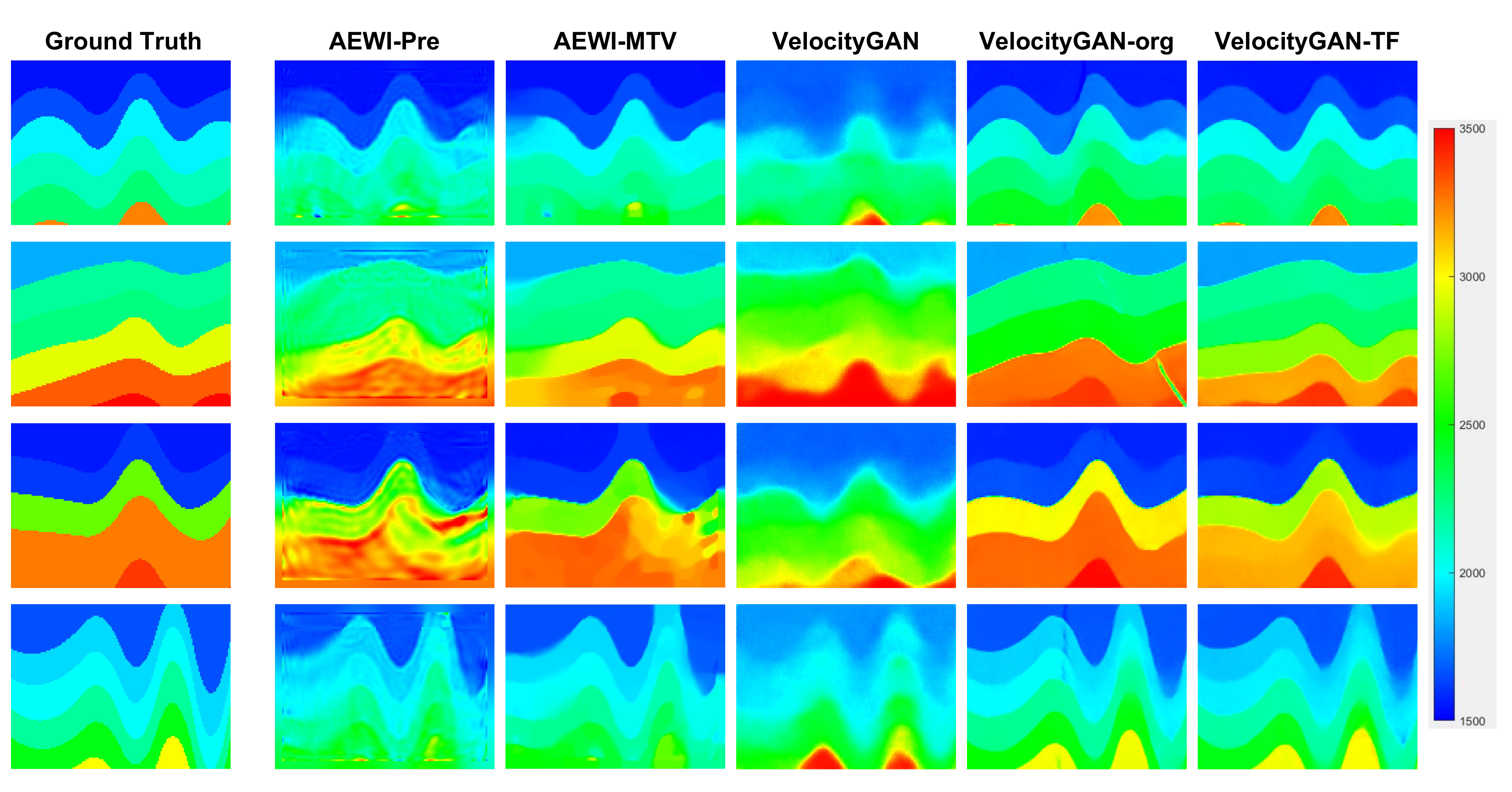}
\end{center}
   \caption{Examples of different methods on 0-Fault CurvedData. The images from left most columns to right are ground truth, reconstruction results using AEWI-Pre, AEWI-MTV, VelocityGAN, VelocityGAN-org, and VelocityGAN-TF.}
\label{fig9: visualization_nofaultData}
\end{figure*}

Besides, we compare the visual appearance of our models on 0-Fault CurvedData in Fig. \ref{fig9: visualization_nofaultData}. Consistent with the quantitative results, VelocityGAN which is totally trained on 0-Fault CurvedData performs worst among these models. It is because that the size of 0-Fault CurvedData is not enough to train a good deep neural network. However, the visual appearance of VelocityGAN-org and VelocityGAN-TF is much better than AEWI-Pre and AEWI-MTV. For physics-driven approaches, especially for AEWI-Pre, there are many oscillations in deep region and high-velocity areas, which can be observed in the second and third rows as shown in Fig.~\ref{fig9: visualization_nofaultData}. The geological interfaces and faults which are generated by data-driven methods are cleaner and sharper. It may because we only calculate the average loss of four selected images for physics-driven methods, the quantitative results can be mis-leading. Compared with VelocityGAN-TF, we can clearly observe that VelocityGAN-org still contains parts of geological faults. The comparison further substantiate that the transfer learning strategy can finetune the deep neural network effectively.

\begin{figure*}[ht]
\begin{center}
    \includegraphics[width=0.9\linewidth]{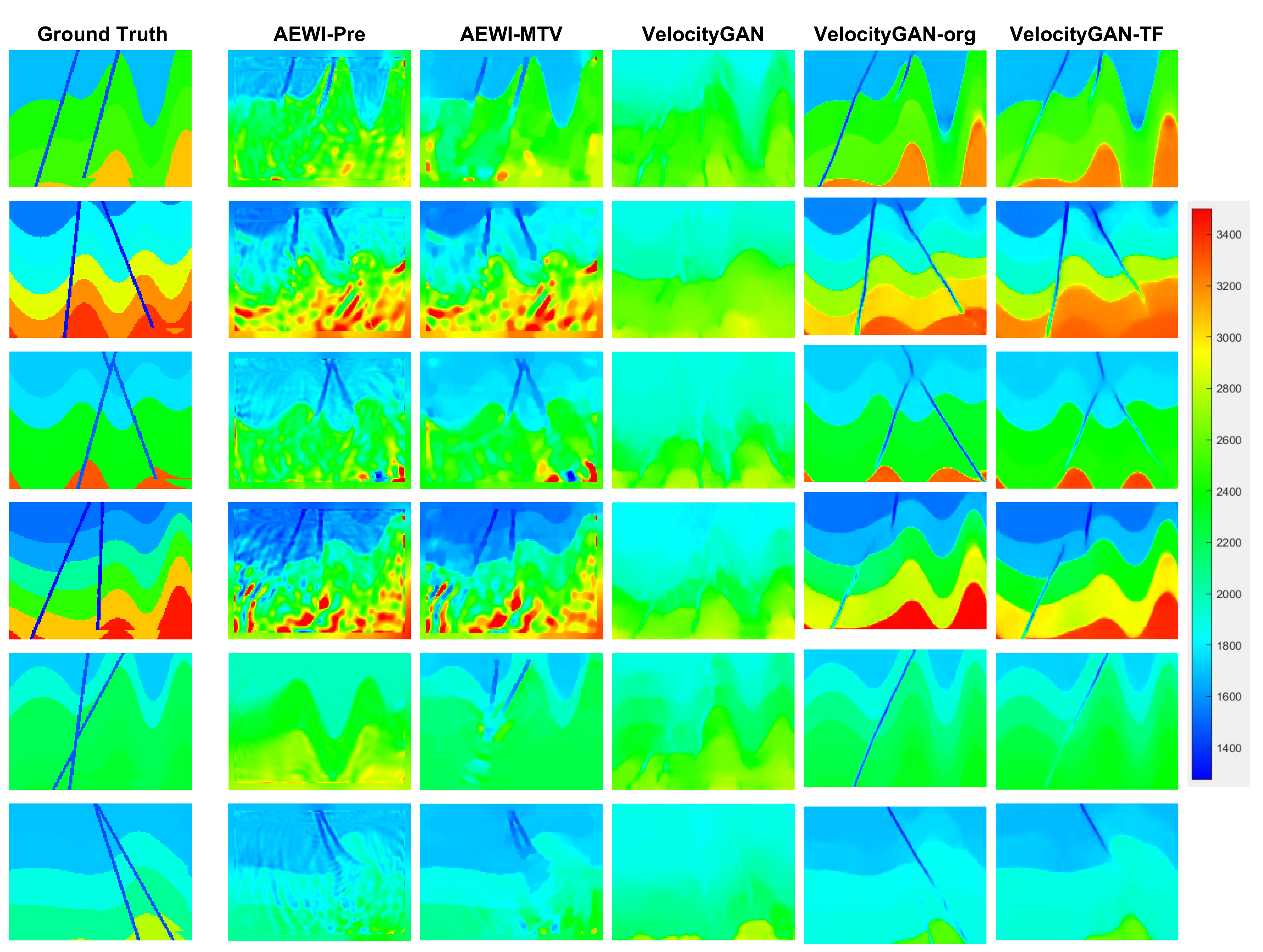}
\end{center}
   \caption{Examples of different methods on 2-Fault CurvedData. The images from left most columns to right are ground truth, reconstruction results using AEWI-Pre, AEWI-MTV, VelocityGAN, VelocityGAN-org, and VelocityGAN-TF.}
\label{fig10: visualization_twofaultData}
\end{figure*}

We randomly select 6 pairs of velocity images from 2-Fault CurvedData and present their reconstruction results in Fig. \ref{fig10: visualization_twofaultData}. As the figure shows, VelocityGAN-TF performs slightly better than VelocityGAN-org, while much better than the other baselines including AEWI-MTV. In both 0-Fault CurvedData and 2-Fault CurvedData, VelocityGAN-org demonstrates its generalization ability to some extent. For example, when the distance between two faults is relatively large (the first four row of Fig. \ref{fig10: visualization_twofaultData}), VelocityGAN-org and VelocityGAN-TF are able to correctly locate two faults, though some of them are fuzzy and unclear. When the distance is relatively small (the fifth and sixth row of Fig. \ref{fig10: visualization_twofaultData}), VelocityGAN-org and VelocityGAN-TF cannot reconstruct correct faults. In this condition, physics-driven approaches, AEWI-Pre and AEWI-MTV, do not perform well either.

To summarize the experiments on 0-Fault CurvedData and 2-Fault CurvedData, we conclude that our VelocityGAN has generalization ability to some extent. For instance, VelocityGAN which is trained only on 1 fault velocity images can also output velocity images which have 0 or 2 faults. With transfer learning strategy, VelocityGAN can further improve its generalization effect, which is competitive with physics-driven methods.

\section{Conclusion}
We develop a data-driven method and we call it ``VelocityGAN,''  to solve the seismic full-waveform inversion problem. We design a modified encoder-decoder structure as the core block of image-to-image target. Based on the encoder-decoder structure, conditional adversarial training strategy with improved loss function are applied to further boost the reconstruction of velocity images. Compared with physics-driven methods, VelocityGAN is a more promising tool for subsurface velocity estimation, because it can alleviate the local minima and expensive computational cost issues. We conduct quantitative and qualitative experiments to demonstrate the effectiveness and efficiency of our VelocityGAN from various aspects. The results substantiate that our model outperforms both the physics-driven methods and the selected deep learning baselines. Furthermore, we also provide extensive experiments to discuss the generalization effectiveness of VelocityGAN. According to the results, we conclude that VelocityGAN has the basic generalization ability and can be improved by transfer learning strategy.

\section*{Acknowledgment}
This work was supported by the Center for Space and Earth Science~(CSES) at Los Alamos National Laboratory (LANL). The computation was performed using super-computers of LANL's Institutional Computing Program.


\bibliographystyle{bst}
\bibliography{main}

\end{document}